# Direct Cortical Control of Primate Whole-Body Navigation in a Mobile Robotic Wheelchair

Sankaranarayani Rajangam[1,2,†], Po-He Tseng[1,2,†], Allen Yin[2,3],
Mikhail A. Lebedev[1,2], Miguel A. L. Nicolelis[1,2,3,4]*

[1] Department of Neurobiology, Duke University Medical Center, Durham, NC.
[2] Duke Center for Neuroengineering, Duke University, Durham, NC.
[3] Biomedical Engineering, Duke University, Durham, NC.
[4] Edmond and Lily Safra International Institute of Neuroscience of Natal, Natal, Brazil.

* Corresponding author: Miguel A. L. Nicolelis, email: nicoleli@neuro.duke.edu
† The two authors contributed equally for this study

**We and others have previously developed brain-machine-interfaces (BMIs), which allowed ensembles of cortical neurons to control artificial limbs (_1-4_). However, it is unclear whether cortical ensembles could operate a BMI for whole-body navigation. Here we show that rhesus monkeys can learn to navigate a robotic wheelchair while seated on top of it, and using their cortical activity as the robot control signal. Two monkeys were chronically implanted with multichannel electrode arrays which simultaneously sampled activity of roughly 150 premotor and sensorimotor cortex neurons per monkey. This neuronal ensemble activity was transformed by a linear decoder into the robotic wheelchair's translational and rotational velocities. During several weeks of training, monkeys significantly improved their ability to navigate the wheelchair toward the location of a food reward. The navigation was enacted by ensemble modulations attuned to the whole-body displacements, and also to the distance to the food location. These results demonstrate that intracranial BMIs could restore whole-body mobility to severely paralyzed patients in the future.**

## INTRODUCTION

The wheelchair remains the main device to assist navigation in people with walking disabilities, particularly those suffering from severe cases of body paralysis (_5_). Up to date, only noninvasive BMI approaches, mostly based on electroencephalography (EEG), have been utilized to enable direct brain control over navigation in a powered wheelchair (_6-9_). Although invasive BMIs hold promise to offer a superior performance over noninvasive systems (_10_, _11_) and approximately 70% of paralyzed patients are willing to accept surgically implanted electrodes in their brains to gain control over their assistive devices (_12_, _13_), invasive BMIs have not been applied yet to wheelchair control.

As a first step towards the development of a clinically relevant device, we utilized large scale recordings from multiple cortical areas (_11_) and our recently developed multichannel wireless recording system (_14_) to enable BMI control over whole-body navigation in a robotic wheelchair.

## RESULTS

The study was conducted in two monkeys (K and M) chronically implanted with multielectrode arrays in multiple cortical areas in both hemispheres (Fig. 1b). Neuronal ensemble activity was sampled using a 512-channel wireless recording system (_14_). In monkey K, 79 neurons were recorded bilaterally in the primary motor cortex (M1), 35 in the right primary somatosensory cortex (S1), and 26 from right dorsal premotor cortex (PMd). In monkey M, 72 neurons were recorded from bilateral M1, and 72 from bilateral S1.

Monkeys were seated in a mobile robotic wheelchair (Fig. 1a; Fig. S1). Each trial started with the robot being placed in one of three starting locations. A food reward (grape) was then dispensed onto a plate mounted at the target location, 1.9 - 2.1 m away. The food location was constant for all trials. Monkeys navigated from the starting location to the food, which completed the trial. At the beginning of each recording session we ran passive navigation trials, with the robot's routes preprogrammed and identical day to day (Fig S2). Cortical neuronal responses to passive whole-body displacements were utilized to obtain initial settings for two Wiener filters (_15_, _16_), which decoded translational ($R$ of prediction of 0.62±0.02 and 0.43±0.02 for monkeys K and M, respectively; mean ± standard error) and rotational (0.30±0.02 and 0.22±0.02) velocity from neuronal ensemble activity (Fig. S3). After the decoders were trained this way, the mode of operation was switched to a direct brain-control BMI navigation mode, where the outputs of the Wiener filters controlled the robot in real time. As we show below, monkeys were able to learn how to use their brain-derived signals to drive this BMI even though the decoding of robot velocity was not superb during the passive navigation training period.

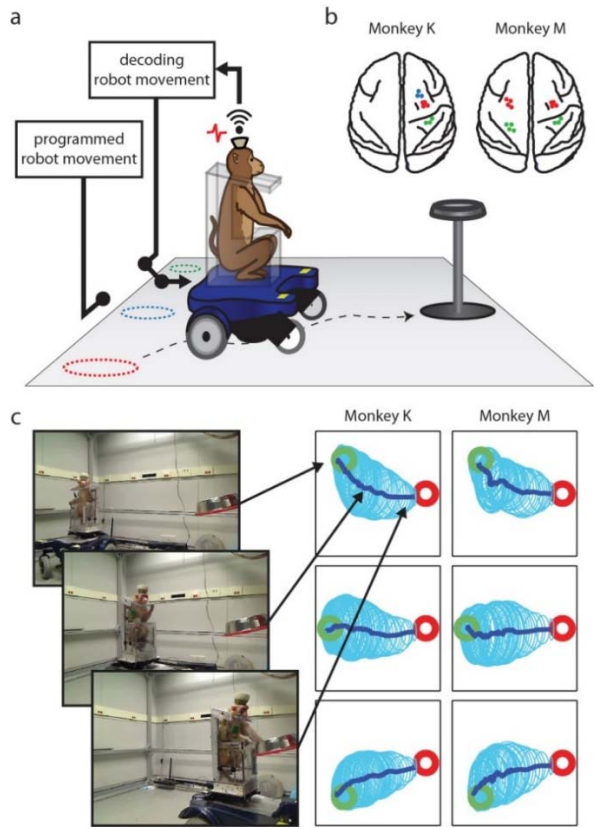

**Fig. 1.** (a) Overview of the experimental design. The mobile robotic wheelchair, which seats a monkey, was moved from one of the three starting locations (dashed circles) to a grape dispenser. The wireless recording system records the spiking activities from the monkey's head stage, and sends the activities to the wireless receiver to decode the wheelchair movement. (b) Schematic of the brain regions from which we recorded units tuned to either velocity or steering. Red dots correspond to units in M1, blue from PMd and green from sensory cortex. (c) Three video frames show Monkey K drive toward the grape dispenser. The right panel shows the average driving trajectories (dark blue) from the three different starting locations (green circle) to the grape dispenser (red circle). The light blue ellipses are the standard deviation of the trajectories.

Individual neurons were tuned to the robot movements during both passive and BMI navigation (*Tuning Index* of 0.24±0.005 and 0.18±0.002 for passive movements, for monkeys K and M, respectively; and 0.15±0.005 and 0.12±0.003 for BMI control; see Materials and Methods). Color plots of Fig. 2a show tuning patterns to translational and rotational velocity for two representative neurons (neurons A and B), both recorded in monkey K's M1. During passive navigation, neuron A increased its firing rate when the robot moved backward, whereas neuron B increased its firing rate when the robot moved backward and/or rotated clockwise. During direct brain-control BMI-based navigation, neuron A remained tuned to backward movement, whereas neuron B was no longer tuned to backward movements, but remained tuned to clockwise rotations. Following our previously introduced approach (*18*), we analyzed tuning depth as a function of time lag between neuronal activity and robot movement for the entire sample of recorded neurons in the 2 monkeys (Fig.2b). This analysis revealed that tuning depth reached its maximum earlier during BMI control than during passive navigation. This result was observed for the entire neuronal population recorded in the two monkeys (Fig. 2c, Wilcoxon signed rank test, p<0.01 for both monkeys) and for M1 and S1 populations analyzed separately (Wilcoxon signed rank test, p<0.01 for both monkeys and both regions). This finding likely indicates the difference in causal relationships between the two modes of operation: during passive navigation neuronal responses were caused by robot movements, whereas during BMI navigation robot movements were caused by neuronal modulations that preceded the actual robotic wheel chair movement. Additionally, maximum tuning depth occurred earlier for M1 neurons than for S1 neurons during BMI control (Wilcoxon Rank Sum Test, p<0.01 for Monkey K (-489.5 ms in M1 vs -431.6 ms in S1); p=0.055 for Monkey M (-495.7 ms in M1 vs -441.3 ms in S1)), but not during passive movements (Wilcoxon Rank Sum Test, p>0.314). This result agrees with previous studies that demonstrated a lead of M1 activity compared to S1 during voluntary movements (*17*), but differs in that movements were initiated through a BMI in our study.

Consistent with our previous studies (*1, 18*) characteristics of neuronal tuning to the robot movements differed between passive navigation and BMI control. Although tuning depth was positively correlated when passive navigation was compared with BMI control across all neurons (Fig. 2d. Mann-Kendall Test, p<0.01), there was a particularly prominent mismatch between the tuning depths for neurons that were weakly tuned (Fig. 2e). This mismatch was particularly strong for monkey M, where it was clear even for the well-tuned neurons (Mann-Kendall Test, p=0.126). In monkey K, well-tuned neurons tended to retain their tuning properties after the transition to BMI control (Mann-Kendall Test, p<0.01).

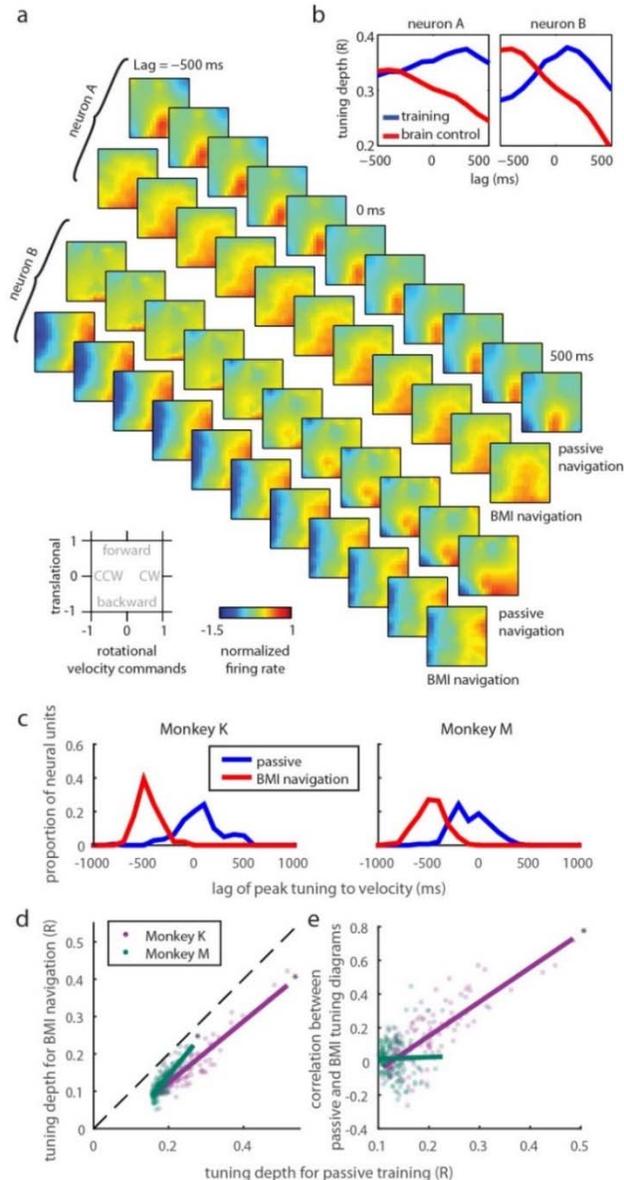

**Fig. 2.** Translational and rotational velocity tuning for neurons tuned during passive training and brain navigation. (a) Each diagram shows the normalized firing rate as a function of translational and rotational velocity commands (bottom-left), where 1 represents the maximum command value sent to the wheelchair (positive values represent forward commands for translation, and clock-wise (CW) for rotation). Each sequence shows the normalized firing rate at different time lag. Neuron A is tuned to backward movement during both passive training and brain control. Neuron B is tuned to right turn movement during brain control, but with different tuning during passive training. (b) The tuning index across time for the two neurons in (a). Note that the time lag of the peak tuning differs between passive training and brain control. (c) Distribution of all the neural units in their tuning index to velocity and the lag of peak velocity tuning. The lag of peak velocity tuning moved from ~0ms to ~500ms earlier than the actions. (d) For neural units that were well tuned during passive navigation, they were more likely to be well tuned during BMI navigation. Each dot represents one neural unit, and the solid line is the regression line. (e) For units that were better tuned, their tuning diagrams between passive and MI navigation were also better correlated in Monkey K, but not in Monkey M.

As the training continued for 3 weeks in monkey M and 6 weeks in monkey K, BMI navigation gradually improved for both monkeys, as evident from the reduction in trial duration (Fig. 3a, Mann-Kendall Test, p=0.025 and 0.045 for Monkey K and M) and trajectory length (Fig.2a, Mann-Kendall Test, p=0.008 and 0.020 for Monkey K and M). Monkey K started with a 43.1 s trial duration (median value for the first week), and improved it to 27.3 s (last week). Monkey M improved from 49.1 to 34.7 s. These performance improvements were accompanied by subtle changes in neuronal tuning. We observed that the decoder settings gradually changed from session to session. Taking the last recording session as a reference, the earlier the session, the less similar were the decoder settings to the last day (Fig. 3b, Mann-Kendal Test, Monkey K: p=0.256 and 0.002 for translational and rotational velocity commands; Monkey M: p=0.032 and p=0.042 for the two commands). This suggests a convergence in changes of neuronal tuning.

After both monkeys reached their best performance levels, we tested whether this improvement resulted from directionally tuned neuronal activity or from non-directional factors (e.g. temporal patterns of neuronal modulations). To change the BMI directional output without changing nondirectional components, we simply reversed the sign of both translational and rotational velocities (i.e., forward instead of backward movement and counterclockwise rotation instead of clockwise). After this operation, the navigation accuracy suffered significant decrease for both monkeys (Fig. 3c, Wilcoxon signed rank test, p<0.01 in both trial duration and trajectory length for both monkeys). This observation confirmed that the monkeys indeed learned to use their directionally tuned cortical neurons to navigate the robot.

Although BMI decoding did not incorporate absolute position of the robot as a controlled parameter, we observed the emergence of positional tuning in cortical neurons during learning of the BMI navigation task. This positional tuning became apparent in the plots that represent neuronal activity as a function of the distance from the food dispenser (Fig. 4). The positional dependence was specific to BMI control: no such dependence was observed during passive navigation (Fig.4a), but it emerged during BMI navigation in both monkeys (Fig.4b, Mann-Kendall trend test, p<0.01 for both monkeys). In monkey K, neuronal rates gradually increased as the robot approached the feeder (Fig.4). In monkey M, there was no such increase until the robot was 0.9 m from the feeder. At this point, neuronal rates started to increase as the robot approached the feeder (Fig.4). These neuronal changes cannot be explained by the monkeys reaching to the food, as the monkeys did not produce arm movements before the robot arrived at the food dispenser location (Video S1). Arm reaching did produce neuronal modulations (highlighted in red in Fig.4), but the comparison of neuronal modulations during arm movements with those during navigation showed only a weak correlation (Fig. S4), which provides support for a different origin of the navigation related cortical modulations than simply arm movements.

## DISCUSSION

In conclusion, our study demonstrated for the first time that cortical neuronal ensembles can directly control whole-body navigation in a mobile device such as a robotic wheelchair. Previously, primate invasive BMI research examined mostly isolated limb movements, whereas neuronal representation of whole-body translations remained little explored. Neuronal mechanisms of spatial encoding in rodents have received more attention (_19_, _20_) but these results are not directly applicable to primates. Until recently, there have been only a few primate studies on spatial representation of the environment by hippocampal neurons (_21_, _22_). A new trend may have started in 2014 as Xu et al. trained monkeys to steer a wheelchair using a hand-held joystick for steering (_23_). They also demonstrated a BMI control of such navigation, where hand movements were decoded from M1 ensemble activity to produce steering commands. They, however, did not attempt to translate cortical activity into whole-body navigation directly, without using hand movements as an intermediary. In another study of joystick driven navigation, monkeys navigated in a virtual environment using a joystick while their bodies remained stationary (_24_). While the joystick control is a viable approach for studies in normal subjects, it is not applicable to severely paralyzed subjects who, like quadriplegics, cannot utilize overt movements to train a BMI decoder. Therefore our direct BMI control of navigation has a practical significance in addition to fundamental questions regarding the representation of whole body movements by cortical neuronal ensembles, a relevant neurophysiological finding that deserves to be further explored.

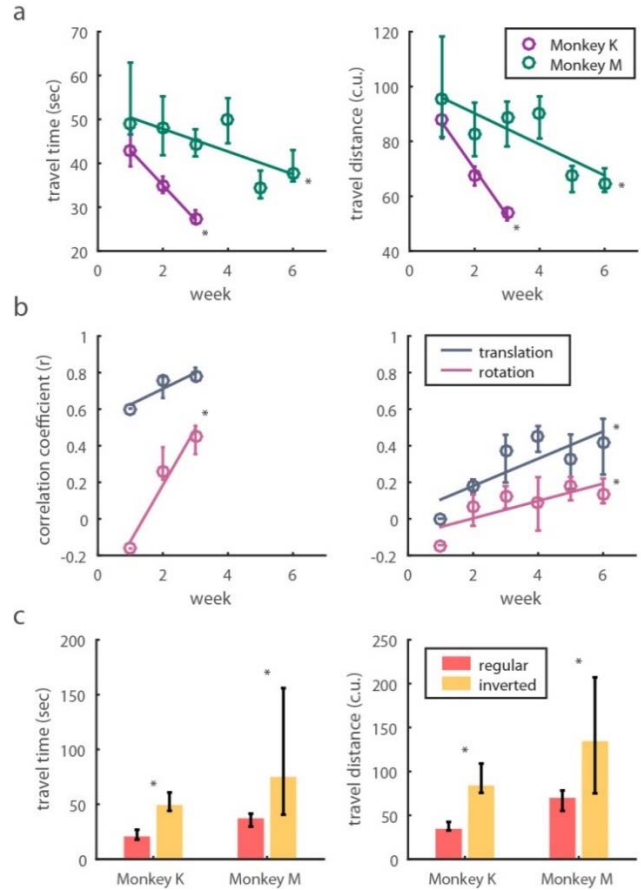

**Fig. 3.** Behavioral improvement and the increase of decoder similarities across sessions. (a) Both monkeys show significant improvement in the traveling time and distance (c.u., camera units) as they learn. The circles represent the median and the error bars show the interquartile range of the medians. (b) Increased correlations between decoders trained in earlier sessions and the last session. (c) Both monkeys demonstrate impaired performance once their decoded movement commands were inverted (forward now becomes backward, and right turn becomes left turn). The bar graph shows the median and the error bars indicate the interquartile range of the medians.

Our approach to enabling BMI control was rather simple: we used passive navigation to set up the decoder parameters, and then relied on monkey cortical plasticity to produce behavioral improvements.

This training sequence did not require any overt movements (*15*, *18*, *23*, *25-27*), an approach which is similar to learning from passive observations that we previously used in our BMIs for uni- and bimanual arm movements (*1*, *3*, *4*). Apparently, passive movements of the entire body evoke somatosensory sensations unlike those earlier studies which were based solely on the vision of a virtual body part moving. Accordingly, it is plausible that proprioceptive and vestibular signals, as well as visual information contributed to the sensorimotor cortex modulations during whole body movement. The loss of somatosensory sensations in the paralyzed may introduce difficulties in the implementation of such a BMI for whole-body navigation. This issue needs more research. Furthermore, the possibility of whether prolonged BMI control of whole-body navigation could result in an incorporation of the navigation device in the brain representation should be explored in future studies. Evidence of such incorporation has been reported even for manually controlled wheelchairs (*28-30*). Our current observations of neuronal plasticity after the transition to BMI control is consistent with such an incorporation as well.

Previously we argued that neuronal ensemble recordings would be utilized in clinically relevant BMIs in the future (*10*). Currently, practical solutions are offered by noninvasive BMIs. Several noninvasive BMIs have been developed to control a wheelchair, such as EEG-based BMIs that utilize motor imagery (*31*) and P300 potentials (*32*). These systems perform with an acceptable (80%) success rate in tasks that involve predefined paths and locations, and can be improved to cope with real environments (*33*). Additional improvements are offered by shared control schemes that incorporate advanced robotics (*34*, *35*). Looking into the future, EEG based systems will probably remain the dominant BMI approach for many years to come, and they will be constantly improving to cover new applications (for example, EEG controlled exoskeletons). However, as invasive systems improve in efficiency and safety, they will become more attractive to the clinical world. Wheelchair navigation using an intracranial BMI could become a real option to severely paralyzed patients in the future.

## MATERIALS AND METHODS

### Study Design

The objective of this study was to demonstrate whole-body navigation with control signals derived from neuronal ensemble recordings in multiple cortical areas. Essential to our design, the learning of BMI driven navigation did not require participants to map overt movements to the navigational direction, which makes this paradigm applicable to the needs of severely paralyzed patients who desire to restore whole-body mobility.

Two adult rhesus macaques (Monkey K and M) were used for the study. All animal procedures were performed in accordance with the National Research Council's Guide for the Care and Use of Laboratory Animals and were approved by the Duke University Institutional Animal Care and Use Committee. The two monkeys were chronically implanted with arrays of microwires in multiple cortical areas of both hemispheres (M1, S1, and PMd). Our recently developed multi-channel wireless recording system was employed to sample from hundreds of neurons in sensorimotor cortex simultaneously (*14*). Both monkeys learned to navigate in a room while seated in a mobile robotic wheelchair and using their cortical activity as the navigation control signal. Cortical ensemble recordings were converted to steering commands for the mobile robotic wheelchair based on linear decoding algorithms.

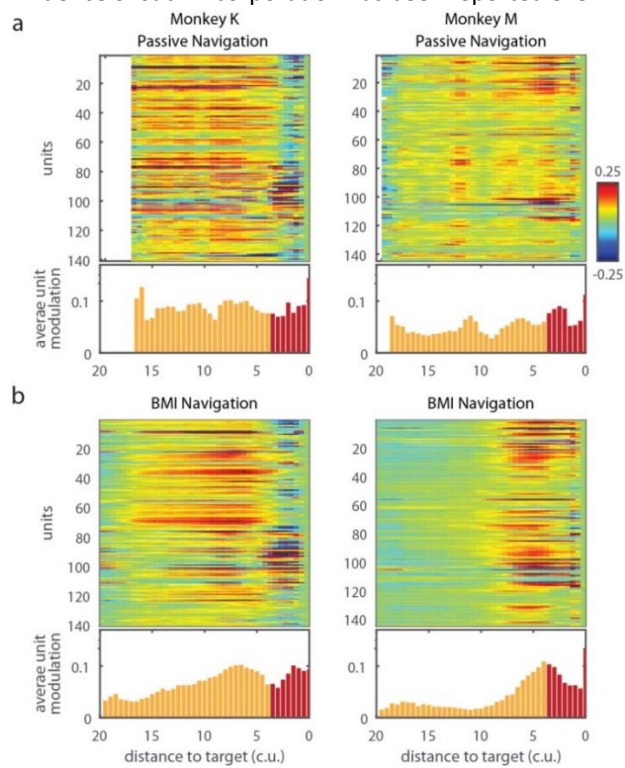

**Fig. 4.** Population responses in Monkey K and M as a function of distance to the target during passive navigation (a) and BMI navigation (b). Color represents normalized firing rate, and the bar graph shows the average unit modulation, calculated by the average of the absolute values of all units' normalized firing rate at each distance (c.u., camera units). When the cart is getting close to the target, the monkeys may reach to the grape, and these reach-related activities are shaded in red.

Both monkeys successfully acquired the ability to steer the robotic wheelchair towards a grape reward. They achieved two dimensional navigation with multiple starting positions and orientations of the robot.

**Task Design**

The monkeys were operantly conditioned to drive the mobile robotic wheelchair toward a food reward (grape). The rewards were delivered by an automated grape dispenser. Each experimental session consisted of about 150 trials during which the monkeys navigated to the grape from a starting location, reached for the grape and placed it in the mouth. The robotic wheelchair was then driven away from the grape dispenser to a new starting location.

The first 30 trials were used to train BMI decoders. During these training trials, the monkeys had no control over the movements of the robotic wheelchair. Rather, the experimental control system navigated the wheelchair along several predetermined routes toward the grape dispenser. The data obtained during this passive navigation was used to train L2-norm Wiener Filters, which generated steering commands from a 1 s window divided into ten 100 ms bins to count spikes produced by each neuron in the ensemble. The filtering outputs were calculated as sums of the spike counts multiplied by the filter weights.

Once the decoders were trained, we proceeded to BMI trials during which the robotic wheelchair was steered directly by the cortical signals. The monkeys would navigate themselves toward the grape, while the wheelchair returning to a starting location was the only part performed automatically by the program. (Obviously the monkeys did not want to move away from the food location.)

In the easiest version of the task, the two monkeys performed 1D navigation (8 sessions for Monkey K; 9 sessions for Monkey M). Here, the monkeys controlled only the forward and backward movements of the robotic wheelchair. The main, more challenging task required the monkeys to navigate in 2D. Their cortical activity was converted into both translational (backward or forward) and rotational (leftward or rightward) velocity components.

The translational velocity was limited to -0.28 to 0.28 m/s (negative values for backward movements and positive for forward), and rotational velocity was limited to -46 to 46 degrees/s (negative for leftward, positive for rightward). If the decoded velocity commands exceeded the limit, the command sent to the robotic wheelchair would be set to the limit value. In the 2D task, the robotic wheelchair could be initially placed at one of the 3 starting locations, whose coordinates were (0.62, 0.50), (0.88, 0), and (0.62, -0.50) meter, where (0, 0) was the room center.

The monkeys navigated in a 3.5-by-2.5m experimental room. The actual drivable area was chosen to be 3.1-by-2.4 m, which assured safety. When the robotic wheelchair was at the drivable area boundary, the program would stop the robotic wheelchair if the decoded commands would have moved the robot through the boundary, and would only execute the decoded commands if the robot moved inside the drivable area. Additionally, when the robotic wheelchair got close to the grape dispenser (docking range), the program would take over the control and park the robot automatically to ensure the monkeys could comfortably obtain their rewards.

**Electrode Implantation**

Monkey M was implanted with four multielectrode arrays. Within each array, Isonel coated, stainless steel polyimide insulated microelectrodes were grouped in two four-by-four, uniformly spaced grids each consisting of 16 microwire triplets of different lengths, for a total of 96 microelectrodes. The implants were placed bilaterally in the primary motor (M1) and primary somatosensory cortices (S1). Each hemisphere received two arrays: one for the upper and one for the lower-limb representations (for a total of 384 microelectrodes). In the current experiment we used 256 channels in monkey M to record from neurons in the bilateral arm and leg areas of M1 and S1. Monkey K was implanted bilaterally with six multielectrode arrays containing 96 microwires each. Stainless steel microwires 30-50μm in diameter were grouped into bundles of three with a single leading conical microwire surrounded by two other microwires with cut angled tips (*14*). The arrays were implanted in the arm and leg representation areas of M1 and S1 and also in the bilateral premotor cortices totaling 576 channels. For the current study we recorded from 128 channels in Monkey K which were from bilateral M1 and premotor cortex and right side S1.

**BMI Navigation System Design**

The BMI navigation system had three main components: *(1) experiment control system*, *(2) wireless recording system*, and *(3) robotic wheelchair*. The experiment control system controlled the experimental sequence, performed BMI decoding of neuronal ensemble activity, and handled the wheelchair video tracking. The wireless recording system recorded neuronal ensemble activity from the monkey brain and sent the neuronal data to the experiment control system. The robotic wheelchair accommodated the monkey chair and received driving commands wirelessly from the experiment control system. The three components communicated with each other in a local network.

**Experiment Control System**

The experiment control system controlled the flow of the experiment, which included the control of the experimental sequence, running BMI decoding algorithms that computed navigation commands to the robotic wheelchair, and closed-loop control of the robotic wheelchair based on the video tracking information.

The location and orientation of the robotic wheelchair was tracked using ASUS Xtion camera (640480 pixels, 30 frames/sec) mounted on the ceiling of the experiment room. The video tracking software was written in C++ and built with OpenCV library. The software processed the video stream, and segmented the video frames to determine the robotic wheelchair location. The wheelchair orientation was determined based on the markers located at the front and the back of the robot. The robot position signals were smoothed using a Kalman filter.

**Wireless Recording System**

Extracellular neuronal recordings were obtained using an in-house built wireless recording system described previously (*14*). Briefly, the wireless recording system comprised the following: digitizing headstages, a wireless transceiver, a wireless-to-wired bridge, and client software. Four headstages were attached to the transceiver module for a total of 128 channels per transceiver unit on which spike sorting could be performed. The bridge received incoming radio packets and converted the signal to an Ethernet interface that collected in the client computer which could be visualized with the custom designed software client. We recorded from 128 channels in Monkey K and 256 channels in Monkey M, yielding 140 and 144 neurons from Monkey K (79 units from M1, 35 from S1, and 26 from Premotor) and Monkey M (72 neurons from M1, and 72 from S1). The number of neurons we acquired from a session was 147.0±10.2 for Monkey K and 157.1±23.9 for Monkey M on average.

**Robotic Wheelchair**

Our robotic wheelchair was a modified, commercially available mid-wheel drive model (Sunfire General) manufactured by Drive Medical. It was powered by two 12volt, 36Ah batteries connected in series to obtain 24VDC. The motors were 450 watt units and were coupled to each drive wheel through a 1:21 ratio transmission, controlled individually to result in directional and speed control. The electromagnetic brakes were removed since braking was obtained with a two channel motor controller added to the unit. The stock VR2 70A motor controller and joystick controls were also completely removed. The powered wheel chair weighed approximately 200Kg and provided a stable low center of gravity platform with a cruising range of 38.8Km rolling on 12 inch diameter drive wheels.

A Roboteq VDC2450 dual channel motor controller served as the interface between an onboard Raspberry Pi (RP) and the wheelchair motors. The RP received the computed motor commands through User Datagram Protocol (UDP) from the experiment control system, and the RP sent the commands to the Roboteq controller via serial data bus connection. As a safety feature, the Roboteq was programmed to stop the robotic wheelchair when the communication failed (i.e., the robotic wheelchair did not receive any motor commands for 1 s), or hit obstacles (i.e. the wheels failed to turn as current limit was set to 50 Amps). An emergency manual power disconnect was prominently placed on the vehicle that would disable the power to the wheels should a malfunction occur that requires a complete manual shut down. A secondary 2.4GHz wireless control system also interfaced to the Roboteq controller and was used as a remote manual wireless control to assist in maneuvering the vehicle between experiments when it was not receiving commands from the experiment control system.

**Grape Dispenser**

Black Corinth Champagne wine grapes were selected to be used as a solid fruit reward because of their very small size and consistency. Long duration experiments involving hundreds of reward transactions can be conducted without the animal losing appetite or interest. The grapes were delivered using an in house designed grape dispenser.

A pneumatic dispenser system was designed and built to allow computer control of this reward. The main mechanical parts of the dispenser were a rotating platter disc and stationary mount plate with a single drop chute. The parts were designed using SolidWorks 3D Cad and were CNC milled from solid blocks of aluminum and have an anodized coating for corrosion resistance. Each dispenser consists of 50 chambers of 13mm diameter, 13mm high oriented on a 340cm diameter rotating platter into which the grapes were loaded.

A pneumatic double acting cylinder (Robart model 166) was used to ratchet advance the platter one chamber at a time around an axis exposing the drop chute that directs the grape to a pedestal, presenting the reward. A secondary pneumatic cylinder (Robart model 165) was mounted vertically over the drop chute and was time delay activated to push out any grapes that may stick in the mechanism over the course of the experiments.

The air cylinders were actuated with 100 psi air through directional control valves (Ingersol Rand model P261SS-012-D-M) and equipped with pneumatic speed controls and regulated to permit the pneumatic actions to be tuned for silence and smooth operation. Optoisolators (PVG612) were used to interface the 24volt control pneumatic valve to the experiment control system via a National Instruments data acquisitions unit (NIDAQ). The Optoisolators were mounted onto a custom designed printed circuit board which we use universally for interfacing to various powered actuators in the lab.

**Statistical Analysis**

We ran Monkey K for 21 sessions and Monkey M for 28 sessions, and sometimes two sessions in a day. Two sessions of Monkey K and five sessions of Monkey M were excluded from the analysis because (1) the sessions had less than 10 BMI navigation trials (3 sessions), (2) technical issues occurred during recording (1 session), (3) a decoder was ill-trained and biased toward negative velocity (1 session), or (4) communication between the experiment control system and the Raspberry Pi on the robot failed frequently (2 sessions). On average, each session yielded 49.0±32.0 and 44.4±9.3 BMI navigation trials by Monkey K and M. In the end, 19 sessions from Monkey K and 23 sessions from Monkey M entered the analysis below.

**Tuning Diagram**

To characterize tuning diagrams for each neural unit at different time lags (e.g., Fig. 2a), we first aligned the controlling commands (i.e., translational and rotational velocity commands) of the robotic wheelchair to the neural activities (spike counts of 100ms time bin). Next, the spike counts were normalized by subtracting the mean and dividing by the standard deviation of the spike counts of this unit. The range of the controlling commands were both divided into 18 bins, which formed the two axes of the tuning diagram (18×18 bins). Then, we sorted the normalized spike counts into the diagram based on the corresponding controlling commands, and averaged the normalized spike counts that fell into the same bin. Lastly, missing values of the tuning diagram were filled in and smoothed by the mean of neighbors in both space and time. To quantify how similar two tuning diagrams were, we calculated Pearson correlation between the two diagrams.

**Tuning Index**

To obtain the *Tuning Index* that characterized the representation of velocity by each neuron, the following multiple linear regression was performed

$$n(t) = \sum_{\tau=-10}^{10} \alpha(\tau)V_t(t-\tau) + \beta(\tau)V_r(t-\tau) + \gamma + \epsilon(t)$$

where $n(t)$ is the spike count at time t; $\tau$ is the lag of time bins (each bin was 100 ms); $V_t(t)$ and $V_r(t)$ are the translational and rotational velocity command at time t, $\alpha(\tau)$ and $\beta(\tau)$ are regression coefficients; $\gamma$ is the intercept; and $\epsilon(t)$ is the residual error. Once the regression model was fitted, we calculated the goodness-of-fit ($R^2$), and the *Tuning Index* was the square root of $R^2$. Because the $\tau$ ranged from 1 s before to 1 s after the spike count measurement, it allowed the tuning index to capture any tuning before or after that time.

We also investigated the tuning in finer temporal resolution (i.e., at each time lag). We applied the equation above; however, instead of summing over all the time lag $\tau$, each $\tau$ was considered separately to obtain *Tuning Index* for each $\tau$. Therefore, we could learn when the neuron was best tuned to the velocity commands.

This equation could also be used to calculate whether a neuron was tuned to other behavioral variables by simply replacing $V_t(t)$ and $V_r(t)$ by other kinematics measure, such as acceleration.

**Behavioral Improvement**

The performance on each trial was characterized by trial duration and the length of the trajectory from the starting location to the reward location. For each session, the medians of trial duration and trajectory length were obtained to represent the behavioral performance of the session. Next, Mann-Kendall Trend Test was performed on these medians for all the sessions to see whether a trend was present.

**Comparison between decoders across days**

To compare decoding across different experimental days, we applied a decoder trained on day A to neuronal data recorded on day B, and vice versa, and measured the change in decoding performance in passive navigation trials. More specifically, to compare two decoders A and B trained on passive navigation of sessions A and B, respectively, we started with identifying neurons that were found in both sessions. Next, we normalized the firing rate of each neuron by subtracting its mean and dividing by its standard deviation. Then, decoder A was trained on session A, and decoder B was trained on session B. Decoder A was then applied to session A and B, and decoder B was applied to both sessions. This way, for each session, we had two predictions of velocity during passive navigation: the one obtained using the same day's decoder and the one obtained by a different day's decoder. The comparison of these two predictions, quantified as Pearson correlation, was used as a measure of the decoders' similarity.

**Inverted Decoder**

After the last session, we ran an additional session to see whether the monkeys learned to utilize the directional properties of the decoder. The hypothesis was that if the monkeys utilized BMI directional commands, their performance would have been significantly impaired after the decoder output was inverted. In this inversion procedure, the absolute values of the translational and rotational velocity commands were kept the same, whereas their signs were reversed: forward movement was turned into backward movement, backward into forward, leftward into rightward, and rightward into leftward. In this control experiment, the monkeys went through 30 trials of passive navigation as usual. Then, the monkey controlled the robotic wheelchair the regular way for the first half of BMI navigations (~45 trials). However, during the second half of BMI navigation (~45 trials), the decoders were inverted.

**Neuronal Modulation to Distance to Target**

The firing rates for each session were first binned into 10ms-bins and normalized against the session averages. The maximum range of distance the wheelchair traveled to was divided into 20 bins. For each session, the normalized firing rates of all neurons corresponding to when the monkey was within a particular distance bin during each regime (training vs. brain-control) were averaged to obtain the session average. The resulting session averages were then averaged to obtain the total average. These were plotted in Fig 4's raster plots.

To obtain the histograms from the total session averages, we took the average of the absolute values of all neuron's normalized firing rates at each distance bin. The reach-dependent distance (the bars colored red in Fig. 4) was a conservative estimate of when a monkey may reach the grape dispenser.

## ACKNOWLEDGMENT

We thank G. Lehew for building the experimental setup and the multielectrode arrays, D. Dimitrov and L. Oliveria for conducting neurosurgeries, D.A. Schwarz for wireless recording support, M. Cervera and T. Phillips for experimental support, and S. Halkiotis for administrative support. **Funding:** This work was supported by National Institute of Health (NIH) grant DP1MH099903 awarded to M.A.L.N. The content is solely the responsibility of the authors and does not necessarily represent the official views of NIH. **Author contribution:** S.R., P.T., M.A.L, and M.A.L.N designed the study. S.R. and P.T. performed the experiment. P.T. wrote the wheelchair localization and control software for the custom BMI suite used for the experiment. P.T. and A.Y. analyzed the data. M.A.L, S.R., P.T., A.Y., and M.A.L.N wrote the manuscript. **Competing interests:** The authors declare that they have no competing interests.

# SUPPLEMENTARY MATERIALS

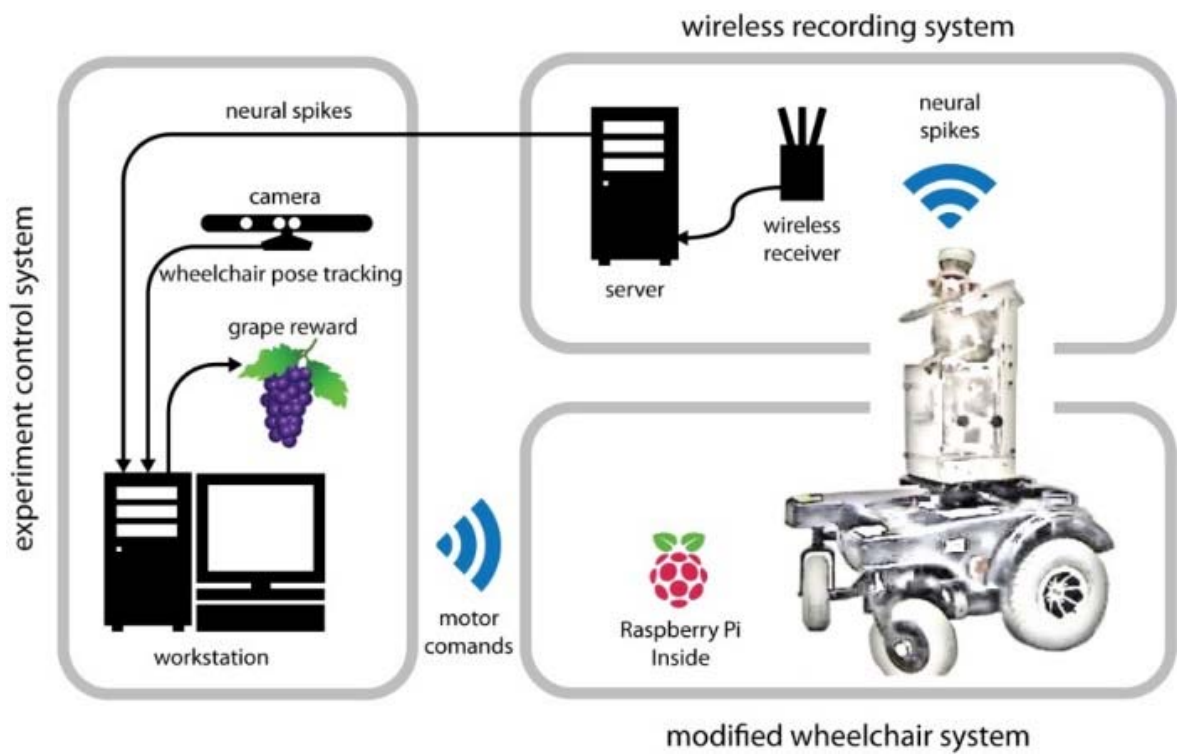

**Fig. S1.** Overview of the BMI navigation system. The driving experiment was composed of three systems. The experiment control system controls the experiment flow, decodes monkey's neural signals, tracks the pose of the wheelchair and delivers the grape reward. The wireless recording system receives the monkey's spiking activities from the monkey's head stage, and sends the activities to the experiment control system. The modified wheelchair system executes the (decoded) wheelchair movement commands from the experiment control

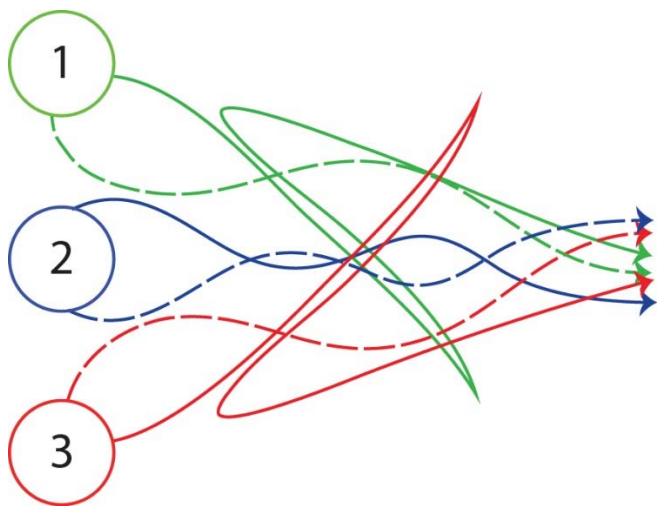

**Fig. S2.** The 6 training routes from three starting locations to the grape dispenser.

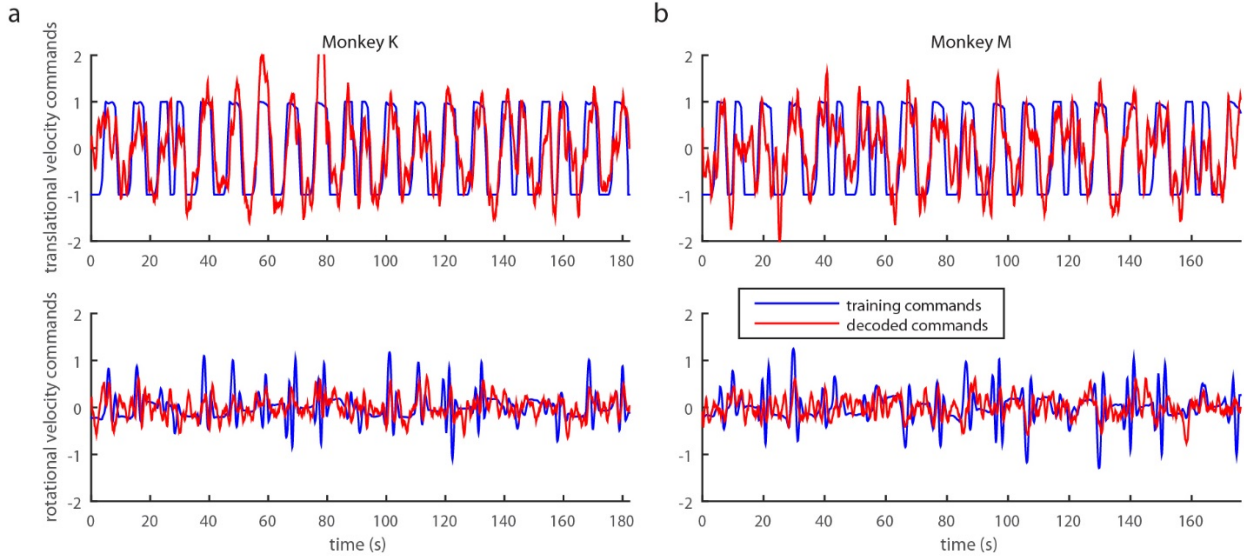

**Fig. S3.** An example of decoded translational and rotational velocity commands during Passive navigation for (a) Monkey K (R = 0.71 and 0.40 for translational and rotational velocity commands) and (b) Monkey M (R = 0.53 and 0.35 for translational and rotational velocity commands). Prediction was evaluated by 5-fold cross validation.

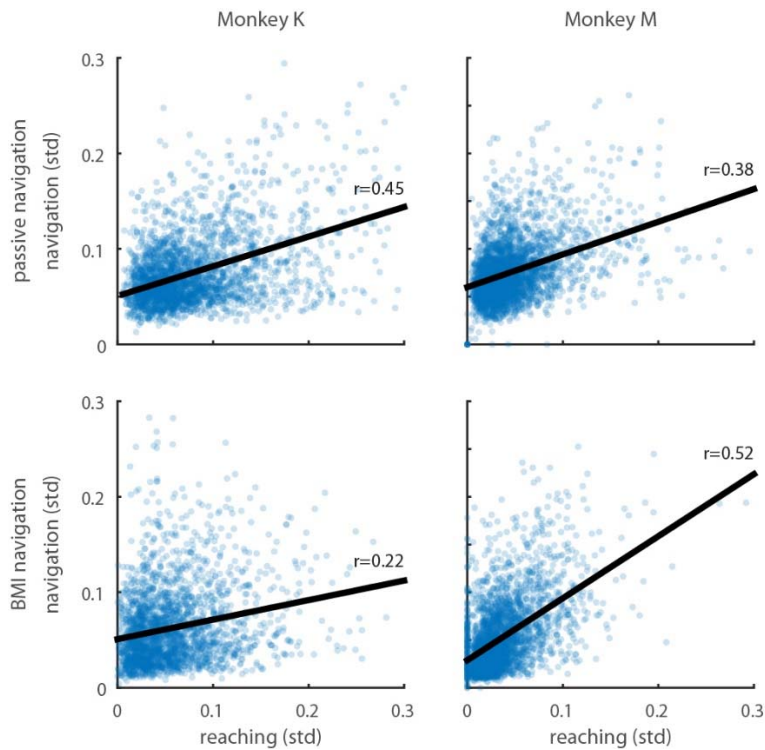

**Fig. S4.** Standard deviations of the normalized firing rates during navigation and reaching. Each dot represents the standard deviation of the normalized firing rate of a neuron in a session. The solid line is the regression line, and the Pearson correlation is shown in the right end of the line (all $p<0.05$).

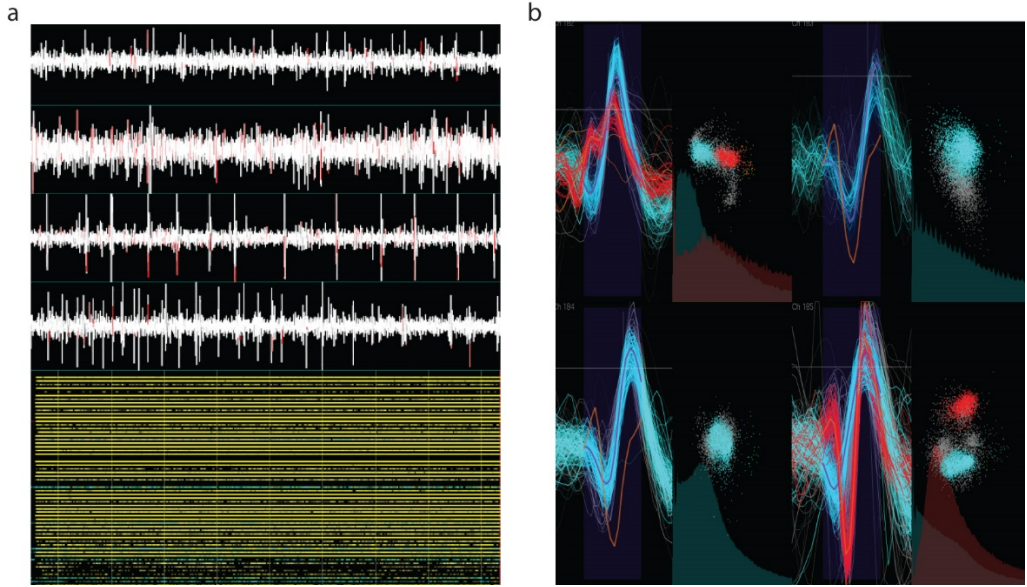

**Fig. S5.** (a) Screenshot of spike voltage traces from four different channels as displayed on the wireless client interface. Raster plot in bottom panel depicts neuronal spiking activity, each row pertains to a single neuron, and each dot is an individual spike, recorded during a session (128 channels displayed) from monkey M. (b) Screenshot of the wireless client with four channels alongside their corresponding PCA clusters and spike templates. Each channel can record a maximum of two units.

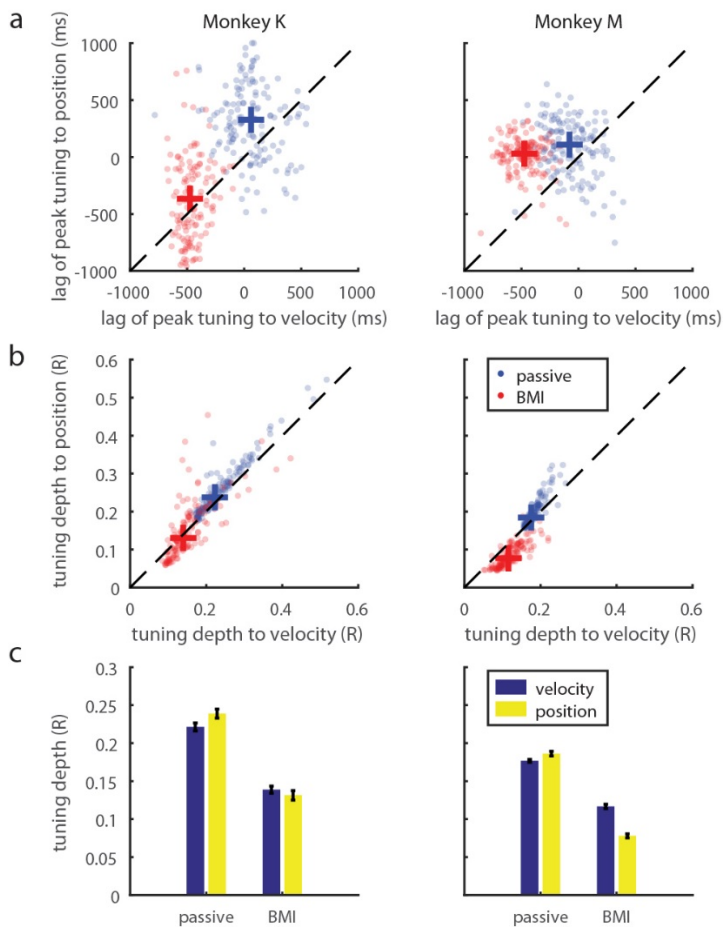

**Fig. S6.** (a) The timing of peak tuning in both velocity (translation and rotation) and position (2D plane) were shifted earlier during BMI navigation than in passive navigation (Wilcoxon signed rank test, p<0.01 for both monkeys). Each dot represents one neural unit under a navigation mode, and the cross indicates the median of all units. (b) The tuning depth (R) decreased when shifted from passive navigation to BMI navigation (Wilcoxon signed rank test, p<0.01 for both monkeys). This result was expected as the decoder used during BMI navigation was trained during passive navigation. The crosses indicate the median of all units, and the medians are also plotted in (c). The error bars represent the upper and lower quartile of the median. While the units were better tuned to position during the passive training, they were better tuned to velocity during brain control (ANOVA, interaction p=0.0675 for Monkey K, and p<0.01 for Monkey M).

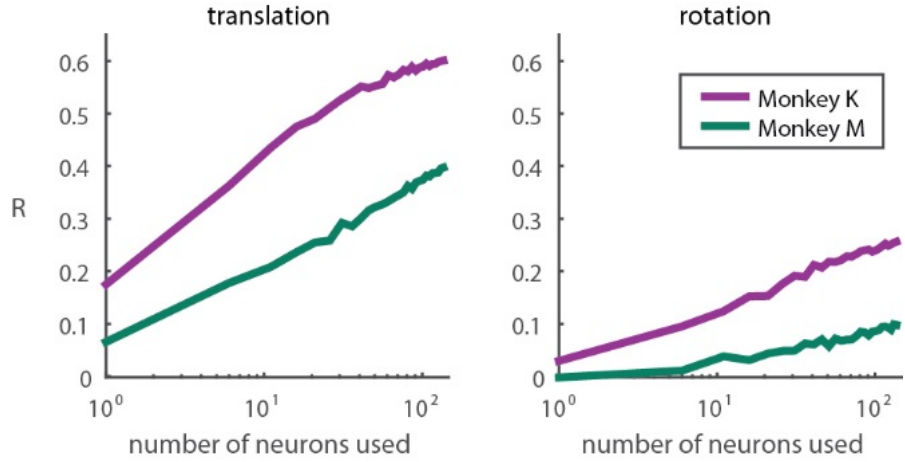

**Fig. S7.** Neuron-dropping curve for decoding translational and rotational velocity commands.

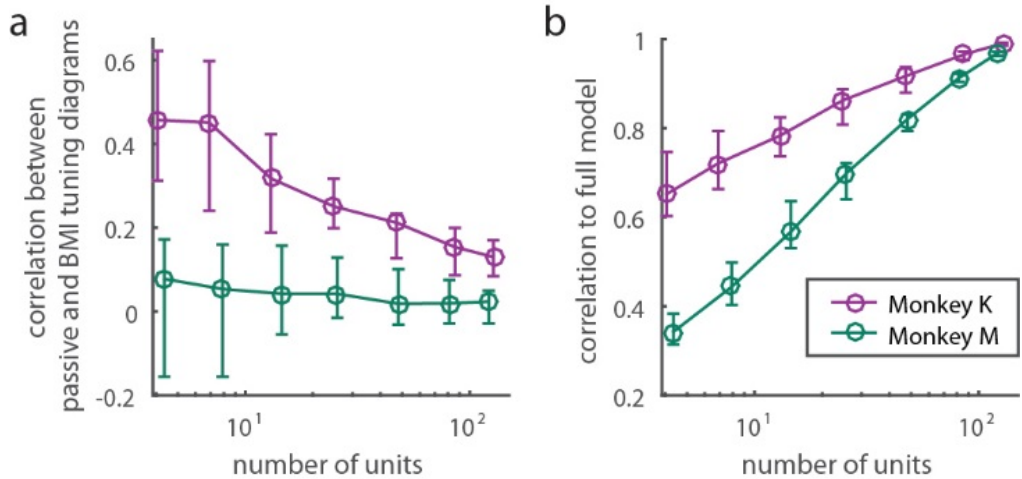

**Fig. S8.** Supplement to Fig. 2e, which shows that a unit's tuning diagrams between passive and BMI navigation were better correlated if the unit is better tuned for Monkey K, but not for Monkey M. This figure shows the same story, but with a group of units that were selected by L1-regularized regression (lasso). Lasso selected a set of units that best decode the translational and rotational velocity commands. (a) Consistent with Fig 2e., Monkey K showed that the best decoding unit set has high correlations in tuning diagram between passive and BMI navigation; the correlation drops when including less informative units (Mann-Kendall Test, $p<0.01$). However, such trend was not observed in Monkey M. (b) The corresponding decoding performance of (a) relative to the full model as a function of number of units.

**Movie S1.** Experiment in a glance

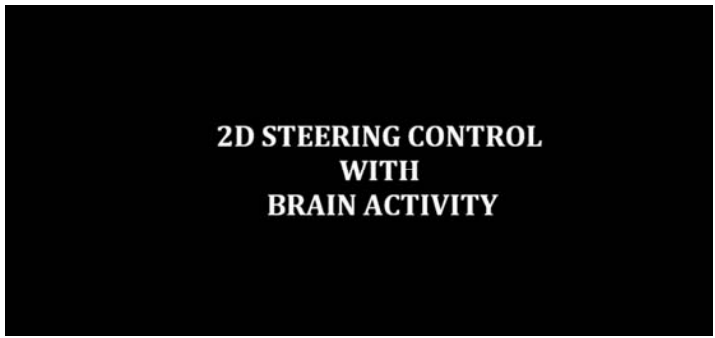

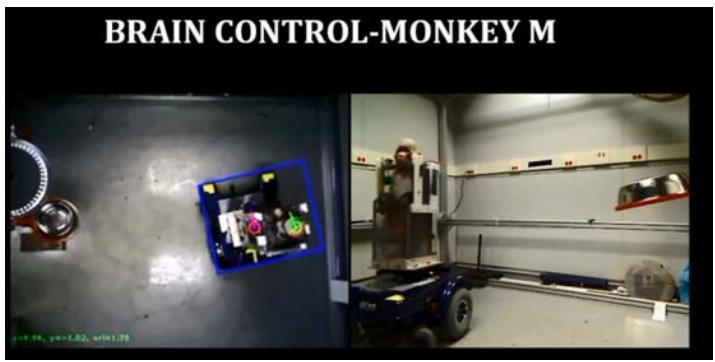

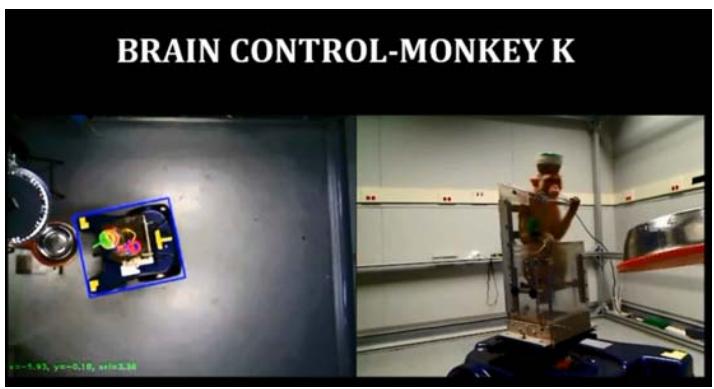